\documentclass[aps,prd,twocolumn,superscriptaddress,nofootinbib]{revtex4-1}

\usepackage{latexsym}
\usepackage{amsmath}
\usepackage{amssymb}
\usepackage{amsfonts}
\usepackage{bm}
\usepackage{physics}

\usepackage{color}
\definecolor{purple}{rgb}{0.5,0,0.5}
\definecolor{blue}{rgb}{0.0,0,0.9}
\definecolor{prdblue}{rgb}{0.133,0.118,0.498}
\usepackage[colorlinks=true, pdfstartview=FitV, linkcolor=prdblue, citecolor= prdblue, urlcolor=prdblue]{hyperref}

\usepackage{supertabular} 
\usepackage{placeins}
\usepackage{epsfig}
\usepackage{graphicx}

\begin{document}


\title{Charge distributions of pseudo-scalar and vector mesons from Dyson-Schwinger equations}


\author{Y.-Z. Xu}
\email[]{yinzhen.xu@dci.uhu.es}
\affiliation{Departmento de Ciencias Integradas, Universidad de Huelva, E-21071 Huelva, Spain.}
\affiliation{Departamento de Sistemas F\'isicos, Qu\'imicos y Naturales, Universidad Pablo de Olavide, E-41013 Sevilla, Spain}

\author{K. Raya}
\email[]{khepani.raya@dci.uhu.es}
\affiliation{Departmento de Ciencias Integradas, Universidad de Huelva, E-21071 Huelva, Spain.}

\author{J. Rodr\'iguez-Quintero}
\email[]{jose.rodriguez@dfaie.uhu.es}
\affiliation{Departmento de Ciencias Integradas, Universidad de Huelva, E-21071 Huelva, Spain.}
\affiliation{Irfu, CEA, Université Paris-Saclay, 91191, Gif-sur-Yvette, France}

\author{J. Segovia}
\email[]{jsegovia@upo.es}
\affiliation{Departamento de Sistemas F\'isicos, Qu\'imicos y Naturales, Universidad Pablo de Olavide, E-41013 Sevilla, Spain}


\date{\today}

\begin{abstract}

We combine the Dyson-Schwinger/Bethe-Salpeter equations framework with modern numerical reconstruction methods to derive the three-dimensional and transverse two-dimensional charge distribution of an array of ground-state pseudoscalar and vector mesons 
from their elastic electromagnetic form factor in the low-momentum region. The charge radii obtained by averaging over the reconstructed charge distributions have been checked to be consistent with those calculated from the slope of the elastic electromagnetic form factor at zero transferred momentum. The capability of the reconstruction procedure for capturing a reliable low-distance charge distribution is discussed and argued to work down to distances of around 0.1 fm, such that it might be potentially applied to extract, {\it e.g.}, mass densities from gravitational form factors.

\end{abstract}


\maketitle


\section{INTRODUCTION}
\label{sec:intro}

The study of meson's fundamental nature constitutes one of the main directions of research in experimental and theoretical hadron physics. The analysis of meson properties such as masses, leptonic decay constants, form factors, etc., should make us better understand their dynamics and its connection with Quantum Chromodynamics (QCD), the quantum field theory for the strong interaction. Particularly, the elastic electromagnetic form factors have attracted wide attention and have been derived within many different calculational schemes and approaches. Among the most successful, the one based on the solutions of the corresponding Dyson-Schwinger/Bethe-Salpeter equations (DSEs/BSEs)\,\cite{Maris:2000sk, Bhagwat:2006pu, Chang:2013nia, Chen:2018rwz, Xu:2019ilh,Hernandez-Pinto:2023yin, Xu:2023vlt,Yao:2024drm} must be underlined; but others also deserve mention as, for instance, lattice-regularized QCD~\cite{Hedditch:2007ex,Koponen:2017fvm,Gao:2021xsm}, dispersive analyses\,\cite{Colangelo:2018mtw} or QCD sum rules~\cite{Aliev:2004uj,Bakulev:2009ib}. On the other hand, apart from the quark model~\cite{Xiao:2002iv, Choi:2004ww}, the earliest and most widely used model to describe the elastic electromagnetic form factor is the so-called single-pole vector meson dominance model (VMD) \cite{Sakurai:1969ss, Fraas:1969uwg}, that can be extended to account for multi-pole structures~\cite{Masjuan:2012sk,Masjuan:2017tvw}. 

We will aim here at a computation of a complete array of elastic electromagnetic form factors for pseudoscalar and vector mesons, and derive from them their associated charge distributions, delivering therewith an intuitive physical picture of their spatial distribution. The usual (non-relativistic) approach is obtaining a meson's or baryon's charge distribution from the three-dimensional (3D) Fourier transform of its electric form factor~\cite{RevModPhys.30.482, Ernst:1960zza, Sachs:1962zzc} 
(their size being therefore expressed by an average squared radius which results from the form factor derivative at zero momentum, as done in PDG compilation\,\cite{ParticleDataGroup:2022pth}). How to extract reliable spatial distributions from hadrons, specially the light ones, is anyhow being under intensive discussions in the last few years\,\cite{Miller:2007uy,Miller:2018ybm,Jaffe:2020ebz,Freese:2021mzg,Epelbaum:2022fjc}. Other transforms are possible (see, \emph{e.g.}, \cite{Jaffe:2020ebz}) which judiciously interpreted should deliver equivalent qualitative pictures\,\cite{Xu:2023izo}. Particularly, one may choose the two-dimensional (2D) transverse charge distribution as a representation for meson's spatial distribution\,\cite{Burkardt:2002hr,Miller:2018ybm}. 

Whether the charge distribution is 2D or 3D, the Fourier transform requires form-factor's data in the full space-like region, which will be herein approached by exploiting the DSE/BSEs formalism, within a non-perturbative and Poincaré-covariant framework capable of simultaneously describing confinement and dynamical chiral symmetry breaking (DCSB), and which has been successfully used to describe a wide range of meson properties~\cite{Roberts:1994dr, Bashir:2012fs,Qin:2020rad,Raya:2024ejx}. That the Bethe-Salpeter amplitudes (BSAs) are usually computed in the hadron's rest frame entails an instrumental difficulty for their direct implementation into the triangle diagram representing the form factor in the impulse approximation. Chebyshev extrapolations were firstly used\,\cite{Maris:2000sk} to circumvent this problem, and some perturbation theory integral representations (PTIR) were more recently applied aiming at the same~\cite{Chang:2013nia, Raya:2015gva, Serna:2020txe}. On the other hand, in Ref.~\cite{Bhagwat:2006pu}, solving BSE based on the moving frame has been proposed, avoiding thus any required extrapolation of the numerical BSEs solutions. A drawback of this approach, owing to the existence of poles in the solution of the quark propagator in complex plane, is its delivering numerically stable outputs for the form factor only restrained to a low momentum region~\cite{Bhagwat:2002tx, Windisch:2016iud, Chen:2018rwz, Jia:2024dfl}.

Consequently, the next step in order to capitalize on the moving-frame BSAs solutions is to propose a well-defined and reliable procedure to extract charge distributions on the basis of a form factor defined within a low-momentum finite region. The latter is a typical ill-posed inversion problem, for which a variety of powerful modern numerical methods have been developed, relying on the recent improvement of computing performance. Among them, the maximum entropy method (MEM) has been recently successfully applied to  analogous problems in hadron physics~\cite{Asakawa:2000tr, Nickel:2006mm, Qin:2014dqa, Gao:2016jka, Xu:2021lxa, Mueller:2010ah, Zhang:2023oja}. Such will be our choice in this work, being herein supplemented with machine-learning techniques for cross validation and errors' estimation, in order to reconstruct both 3D and 2D charge distributions from elastic electromagnetic form factors. 

The manuscript is organized as follows. We introduce the calculation of meson's electric form factor within the DSEs/BSEs framework in Sec.~\ref{sec:FFs-DSEsBSEs}, whereas the reconstruction of the related charge distribution can be found in Sec.~\ref{sec:Distributions}. In Sec.~\ref{sec:Results}, we show the numerical results for the charged states\,\footnote{Herein $\pi_{s,c,b}$ are charged $\pi^+$-like systems, defined by considering a first flavor generation of two quarks with a degenerate mass as $s,c,b,$ quarks, respectively; and $\rho_{s,c,b}$ standing for the analogous charged vector meson counterparts.} $\pi,\pi_{s,c,b}, \rho,\rho_{s,c,b},K,K^*$; and the neutral ones\footnote{We used the non-standard notation $K_0^\ast$ for the first spin excitation of the neutral kaon.} $K_0,K_0^*$, which are therein separately discussed. Finally, Sec.~\ref{sec:Summary} provides a brief summary and some concluding remarks, while further elaborations about the reliability of the reconstruction procedure are presented in two appendices. 


\section{ELECTRIC FORM FACTORS}
\label{sec:FFs-DSEsBSEs}

The DSEs/BSEs framework formulated in Euclidean space and within the rainbow-ladder (R-L) truncation is applied herein. Therefore, the generalized impulse approximation allows us to describe electromagnetic processes in terms of dressed quark propagators, meson's BSAs, and the dressed quark-photon vertex; these couplings can be written as~\cite{Maris:2000sk, Bhagwat:2006pu, Xu:2019ilh}
\begin{align}
\Lambda_H^{\mu, f g g}(P, Q) & =i N_c \int_k \operatorname{Tr}\left[S^f\left(k_p\right) \Gamma_H\left(k_{-} ; k_p\right) S^g\left(k_{+}\right)\right. \nonumber \\
& 
\left.\times \Gamma^\mu\left(k_{+} ; k_{-}\right) S^g\left(k_{-}\right) \Gamma_H\left(k_{-} ; k_p\right)\right] \,,
\label{Eq.triangle}
\end{align}
where
\begin{equation}
k_{+}=k+\frac{P}{2}+\frac{Q}{2}, \quad k_{-}=k+\frac{P}{2}-\frac{Q}{2}, \quad k_p=k-\frac{P}{2},
\end{equation}
and $P-Q/2, P+Q/2, Q$ are incoming meson, outgoing meson and incoming photon momenta, in that order. Note that the on-shell condition fixes the following relation
\begin{equation}
(P-Q/2)^2 = (P+Q/2)^2 = -M^2 \,,
\end{equation}
with $M$ the meson's mass.

The dressed quark propagators $S(\tilde{k})$, meson's BSAs $\Gamma_H(\tilde{k}_+,\tilde{k}_-)$, and the dressed quark-photon vertex $\Gamma^{\mu}(\tilde{k}_+,\tilde{k}_-)$, can be obtained by solving the gap equation,
\begin{equation}
\begin{aligned}
S^{-1}(\tilde{k}) &=Z_2 i \gamma \cdot \tilde{k}+Z_4 m \\
&
+Z_1 \int_{\tilde{q}}^{\Lambda} g^2 D_{\alpha \beta}(\tilde{k}-\tilde{q}) \frac{\lambda^a}{2} \gamma_\alpha S(\tilde{q}) \frac{\lambda^a}{2} \gamma_\beta \,,
\label{eq:DSE}
\end{aligned}
\end{equation}
the homogeneous BSE,
\begin{equation}
\Gamma_H\left(\tilde{k}_{+}, \tilde{k}_{-}\right) = \int_{\tilde{q}}^{\Lambda} K(\tilde{q},\tilde{k};\tilde{P})  S\left(\tilde{q}_{+}\right) \Gamma_H\left(\tilde{q}_{+}, \tilde{q}_{-}\right) S\left(\tilde{q}_{-}\right) \,,
\label{eq:hBSE}
\end{equation}
and the inhomogeneous BSE,
\begin{equation}
\begin{aligned}
\Gamma_\mu\left(\tilde{k}_{+}, \tilde{k}_{-}\right)  & = Z_2 \gamma_\mu \\ &- \int_{\tilde{q}}^{\Lambda}  K(\tilde{q},\tilde{k};\tilde{P})  S\left(\tilde{q}_{+}\right) \Gamma_\mu\left(\tilde{q}_{+}, \tilde{q}_{-}\right) S\left(\tilde{q}_{-}\right) \,,
\label{eq:iBSE}
\end{aligned}
\end{equation}
respectively. Where $\tilde{q}_\pm=\tilde{q}\pm \tilde{P}/2$, $m$ is the current-quark mass and $Z_{1,2,4}$ are the renormalization constants. Here, we employ a mass-independent momentum-subtraction renormalization scheme and choose the associated scale to be $\mu=19\,\text{GeV}$~\cite{Chang:2008ec, Xu:2021mju}, $\int_{\tilde{q}}^{\Lambda}$ represents a translationally-invariant regularization of the four-dimensional integral with the regularization scale $\Lambda$. Under R-L approximation, the scattering kernel can be written as
\begin{equation}
K(\tilde{q},\tilde{k};\tilde{P}) = Z_1g^2 D_{\alpha \beta}(\tilde{k}-\tilde{q}) \frac{\lambda^a}{2} \gamma_\alpha \otimes \frac{\lambda^a}{2} \gamma_\beta \,,
\label{eq:K}
\end{equation}
with
\begin{equation}
Z_1 g^2 D_{\mu \nu}(\tilde{k})=Z_2^2 \mathcal{G}(\tilde{k}^2) \mathcal{P}_{\mu \nu}^T(\tilde{k}) \,,
\end{equation}
and $\mathcal{P}_{\mu \nu}^T(\tilde{k})=\delta_{\mu \nu}-{\tilde{k}_\mu \tilde{k}_v}/\tilde{k}^2$, which is the transverse projection operator. Concerning the effective interaction, $\mathcal{G}(\tilde{k}^2)$, we employ the Qin-Chang model~\cite{Qin:2011xq}:
\begin{equation}
\frac{\mathcal{G}(\tilde{k}^2)}{\tilde{k}^2}=D\frac{8\pi^2}{\omega^4}e^{-\tilde{k}^2/\omega^2}+\frac{8\pi^2\gamma_{m}\mathcal{F}(\tilde{k}^2)}{\ln[\tau+(1+\tilde{k}^2/\Lambda^2_{\text{QCD}})^2]}
\label{eq:g}
\end{equation}
with $\mathcal{F}(\tilde{k}^2)=\left\{1-\exp[(-\tilde{k}^2/(4m_t^2)]\right\}/\tilde{k}^2$, $m_t=0.5$\,GeV, $\tau=e^2-1$, $\Lambda_{\text{QCD}}=0.234$\,GeV, $\gamma_{m}=12/25$. In order to generate sensible values of masses and decay constants, we use typical values of the parameters $\omega$ and $D$, namely $\omega=0.5\ \mathrm{GeV}$ and $D \omega=(0.8\ \mathrm{GeV})^3$ for mesons containing only $u/d$ and $s$ quarks; and, $\omega=0.8\ \mathrm{GeV}$ with $D \omega=(0.6\ \mathrm{GeV})^3$ for heavy quarkonia\,\cite{Bhagwat:2004hn, Xu:2021mju}. More details about solving Eqs.~\eqref{eq:DSE}-\eqref{eq:iBSE} with the model specified in Eqs.~\eqref{eq:K}-\eqref{eq:g} can be found in, for instance, Refs.~\cite{Maris:1997tm, Qin:2011xq}.

\begin{table}[!t]
\caption{\label{tab:mf} Masses and decay constants, in GeV, for pseudo-scalar and vector mesons. The renormalization-group-invariant current-quark mass is $\hat{m}$; $M$ is the meson's mass; $f$ is the decay constant. For comparison, the reported experimental data are~\cite{ParticleDataGroup:2018ovx, Qin:2019oar, McNeile:2012qf}: $M_\pi=0.138(1)$, $f_\pi=0.092(1)$,$M_{\rho}=0.775(1)$, $f_\rho=0.153(1)$, $M_\phi=1.019(1)$, $f_\phi=0.168(1)$, $M_{\eta_c}=2.984(1)$, $f_{\eta_c}=0.237(52)$, $M_{J/\psi}=3.097(1)$, $f_{J/\psi}=0.294(5)$, $M_{\eta_b}=9.399(1)$, $M_{\Upsilon}=9.460(1)$, $f_{\Upsilon}=0.505(4)$.}
\begin{ruledtabular}
\begin{tabular}{clllllll}
quark & \multicolumn{1}{c}{$\hat{m}$} & meson & \multicolumn{1}{c}{$M_{0^-}$} & \multicolumn{1}{c}{$f_{0^-}$} & meson & \multicolumn{1}{c}{$M_{1^-}$} & \multicolumn{1}{c}{$f_{1^-}$} \\
\hline
$u/d$ &0.007 &$\pi_{\ }$& 0.135 & 0.095 &$\rho$ & 0.755 & 0.150\\
$s$ &0.160 &$K$ &0.495 & 0.112 & $K^*$ & 0.955 &0.178\\
\multicolumn{1}{c}{/}    &   \multicolumn{1}{c}{/}    &$\pi_s$& 0.692 & 0.133 & $\rho_s$& 1.088 & 0.189\\
$c$ &1.762 &$\pi_c$& 2.984 & 0.260 &$\rho_c$ &3.094 & 0.274\\
$b$ &7.473 &$\pi_b$& 9.399 & 0.479 & $\rho_b$&9.462 & 0.461
\end{tabular}
\end{ruledtabular}
\end{table}

The novelty of this work is that we solve Eqs.~\eqref{eq:DSE} to \eqref{eq:iBSE} in the moving frame~\cite{Bhagwat:2006pu, Xu:2019ilh}
and then the quark-photon coupling, Eq.~\eqref{Eq.triangle}, can be derived. Note here that the coupling of a photon to the quark and antiquark pair should be written as a sum of two terms,
\begin{equation}
\Lambda_H^\mu(P, Q)=\hat{Q}^g \Lambda_H^{\mu, \bar{f} g g}(P, Q)+\hat{Q}^{\bar{f}} \Lambda_H^{\mu, g \bar{f} \bar{f}}(P, Q) \,,
\label{eq:ff}
\end{equation}
where $\hat{Q}$ is the quark/antiquark electric charge.
 
For pseudo-scalar mesons, the electromagnetic form factor is unique
\begin{equation}
G_E^{\mathrm{PS}}\left(Q^2\right)=F(Q^2)=\frac{P^\mu}{2 P^2} \Lambda^\mu(P, Q) \,;
\end{equation}
whereas, for vector mesons, there are actually three electromagnetic form factors. In this work, we just focus on the electric one:
\begin{align}
G_E^{\mathrm{VC}}\left(Q^2\right) & =\left(1+\frac{2}{3} \frac{Q^2}{4 M^2}\right) F_1\left(Q^2\right)+\frac{2}{3} \frac{Q^2}{4 M^2} F_2\left(Q^2\right) \nonumber \\
& 
+ \frac{2}{3} \frac{Q^2}{4 M^2}\left(1+\frac{Q^2}{4 M^2}\right) F_3\left(Q^2\right),
\end{align}
with 
\begin{equation}
\Lambda_{\rho \sigma}^\mu(P, Q) = -\sum_{j=1}^3 T_{\mu \rho \sigma}^j(P, Q) F_j\left(Q^2\right) \,, 
\end{equation}
and
\begin{subequations}
\begin{align}
T_{\rho \sigma}^{\mu, 1}(P, Q) & =2 P_\mu \mathcal{P}_{\rho \gamma}^T\left(P^{-}\right) \mathcal{P}_{\gamma \sigma}^T\left(P^{+}\right), \\
T_{\rho \sigma}^{\mu, 2}(P, Q) & =\left(Q_\rho-P_\rho^{-} \frac{Q^2}{2 M^2}\right) \mathcal{P}_{\mu \sigma}^T\left(P^{+}\right) \nonumber \\
&
-\left(Q_\sigma+P_\sigma^{+} \frac{Q^2}{2 M^2}\right) \mathcal{P}_{\mu \rho}^T\left(P^{-}\right), \\
T_{\rho \sigma}^{\mu, 3}(P, Q) & = \frac{P_\mu}{M^2}\left(Q_\rho-P_\rho^{-} \frac{Q^2}{2 M^2}\right) \nonumber \\
&
\times \left(Q_\sigma+P_\sigma^{+} \frac{Q^2}{2 M^2}\right).
\end{align}
\end{subequations}
The electric form factor at $Q^2=0$, $G_E(0)$, defines the meson's charge: 
\begin{equation}
G_E^{\mathrm{VC}}(0)=G_E^{\mathrm{PS}}(0)=e,
\end{equation}
and $e=1$ for a meson with unit positive electric charge, $e=0$ for a neutral meson.

It is worth noting that, although this approach avoids any extrapolation and fitting, it comes at the cost of only accurately computing the form factors of mesons in low momentum finite region \cite{Bhagwat:2002tx,Windisch:2016iud,Chen:2018rwz}. In the next section, we will present how to extract the charge distributions from them.


\section{CHARGE DISTRIBUTIONS}
\label{sec:Distributions}

The electric form factors, obtained as discussed in the previous section, and the corresponding charge distributions relate to each other by a Fourier transform, which reads~\cite{RevModPhys.30.482, Ernst:1960zza, Sachs:1962zzc} 
\begin{equation}
F(Q^2)=4\pi\int_0^{\infty} dr\, j_0(Qr) r^2 \rho^{\rm{3D}}(r) \,,
\label{eq:3d}
\end{equation}
where the charge density resulting from the usual non-relativistic approach in three space-like dimensions (3D) is denoted as $\rho^\textrm{3D}$, and $j_0$ stands for the zeroth spherical Bessel function. As discussed in the introduction, some controversy\,\cite{Jaffe:2020ebz,Freese:2021mzg,Epelbaum:2022fjc,Xu:2023izo}  has been recently raised about the interpretation and reliability of non-relativistic objects as 3D or transverse 2D densities, the latter defined for the charge as\,\cite{Miller:2010tz, Miller:2018ybm, Burkardt:2002hr}
\begin{equation}
F(Q^2)=2\pi \int_0^{\infty} db\, J_0(Qb) b \rho^{\rm{2D}}(b) \,,
\label{eq:2d}
\end{equation}
where $J_0$ is the zeroth Bessel function of the first kind. In the vector case, one keeps $F(Q^2)=G_E^{VC}(Q^2)$ and ignores form factor corrections that originate from projecting form factors defined in 3 dimensions, to 2D\,\cite{Kim:2022bia}. An exact relation between charge distribution and form factor is still under investigation\,\cite{Jaffe:2020ebz, Li:2022hyf}. 

Let us first focus on the 3D case and capitalize on the Taylor expansion of $j_0$ to recast Eq.\,\eqref{eq:3d} as
\begin{align}\label{eq:Fmom}
F(Q^2) = \sum_{n=0}^{\infty} (-1)^n \frac{Q^{2n}}{(2n+1)!} \, \langle r^{2n} \rangle \;,
\end{align}
with 
\begin{align}
\label{eq:r2n}
\langle r^{2n} \rangle = 4\pi  \int_0^\infty dr \, r^{2+2n} \rho^\textrm{3D}(r) \;.     
\end{align}
As illustrated in App.\,\ref{app:convergence}, Eq.\,\eqref{eq:Fmom} is only valid provided that $\rho^\textrm{3D}(r)$ decreases faster than any power of $r$ at asymptotically large $r$; and its rhs's series is only convergent within a given range of $Q^2$, depending on the particular density profile. On the other hand, considered as an asymptotic series, the form factor derivatives at vanishing momenta can be directly related to the charge distribution moments, \emph{viz.} Eq.\,\eqref{eq:A1}. In particular, for $n$=1, one is left with the well-known result
\begin{align}\label{eq:3dr2}
\langle r^2 \rangle = \left. - 6\, \frac{d}{dQ^2} F(Q^2)\right|_{Q^2=0} \;.    
\end{align}

Analogous arguments work for the 2D case, such that Eq.\,\eqref{eq:2d} leads to
\begin{subequations}
\begin{align}
\label{eq:Fmomb}
F(Q^2) &= \sum_{n=0}^\infty (-1)^n \frac{Q^{2n}}{2^{2n} (n!)^2} \langle b^{2n} \rangle \;, 
\\
\label{eq:b2n}
\langle b^{2n} \rangle &= 2\pi \int_0^\infty db \, b^{2n+1} \rho^\textrm{2D}(b) \;; 
\end{align}    
\end{subequations}
from which, for $n$=1, one obtains  
\begin{align}\label{eq:2db2}
\langle b^2 \rangle = \left. -4\, \frac{d}{dQ^2} F(Q^2)\right|_{Q^2=0} \,.    
\end{align}

\begin{figure}[!t]
\centering
\includegraphics[trim = 0mm 5mm 0mm 0mm, clip, width=1.0\columnwidth]{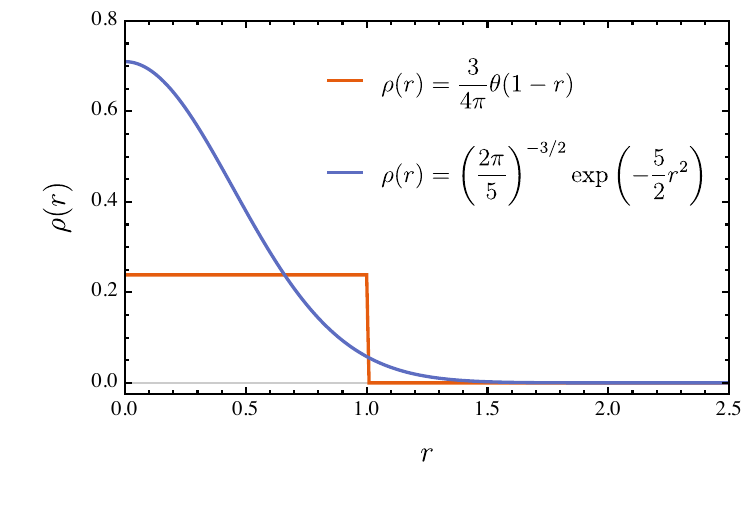}
\caption{\label{fig:test} An example of two very different charge distributions giving the same value of the charge radius, $\langle r^2 \rangle = 3/5$.}
\end{figure} 

It is apparent from Eqs.\,(\ref{eq:Fmom},\ref{eq:Fmomb}) that an accurate knowledge of the form factor, even within a restricted domain of momenta, makes \emph{a priori} available the determination of moments for 2D or 3D distributions, the more accurate the more moments. The zeroth moments are fixed by the normalization condition
\begin{align}\label{eq:norm}
\langle r^0 \rangle = \langle b^0 \rangle = e \;;   
\end{align}
while the first ones, Eqs.\,(\ref{eq:3dr2},\ref{eq:2db2}), define the charge radius and have been widely used to evaluate hadron's sizes~\cite{ParticleDataGroup:2022pth, Carlson:2015jba, Xu:2019ilh, Feng:2019geu}. However, a distribution is only fully defined by the knowledge of all its moments (as a simple illustration, Fig.\,\ref{fig:test} exhibits how different two normalized distributions sharing the same charge radius could be). In practice, a sensible number of moments is needed to fairly approximate the density profile, which offers a more precise physical picture of the hadron than its simple size.  

Alternatively, the charge distribution can be directly obtained by inverting the Fourier transform in Eqs.~\eqref{eq:3d} and \eqref{eq:2d}~\cite{Chang:2009ae}, with the results for the electric form factors as an input. However, specially to derive an output at small $r$, a numerically challenging evaluation of form factors at very large momentum is required. This complication is often bypassed by considering a physically motivated extrapolation\,\cite{Zhang:2021mtn,Raya:2021zrz,Xu:2023bwv,Xu:2023izo} of the form factor.

In this work, we will proceed differently and, as far as the inversion of Eqs.\,(\ref{eq:3d},\ref{eq:2d}) is a typical ill-posed problem, apply a technique denominated maximum entropy method (MEM), which has been proved efficient within this context in hadron physics\,\cite{Asakawa:2000tr, Nickel:2006mm, Qin:2014dqa, Gao:2016jka, Xu:2021lxa, Mueller:2010ah, Zhang:2023oja}. The procedure can be outlined as follows. 

Let us schematically represent Eqs.\,(\ref{eq:3d},\ref{eq:2d}) as 
\begin{equation}\label{eq:FTgen}
F(Q^2)=\int_{0}^{\infty} d\tilde{r}\, \mathcal{K}(Q,\tilde{r})\rho(\tilde{r}) \,,
\end{equation}
where $\mathcal{K}$ is the integral kernel, either $4\pi {\tilde{r}}^2 j_0(Q\tilde{r})$ or $2\pi \tilde{r} J_0(Q\tilde{r})$, with $\tilde{r}=r$ or $b$, respectively, for 3D or 2D. The distribution function $\rho(\tilde{r})$ is assumed to satisfy the normalization condition and the asymptotic property at large $\tilde{r}$ expressed above; and, under MEM framework, can be reconstructed by maximizing the functional
\begin{equation}
\mathcal{Q}[\rho]=\alpha S[\rho]-L[\rho],
\label{eq:MEM}
\end{equation}
where $L[\rho]$ is the likelihood function 
\begin{equation}\label{eq:Lrho}
L[\rho] = \sum_i \frac{1}{2\sigma^2_i} \left[F(Q_i^2) - \int_0^\infty d\tilde{r}\, \mathcal{K}(Q_i,\tilde{r}) \rho(\tilde{r}) \right]^2,
\end{equation}
corresponding to an ordinary $\chi^2$-fitting, $\alpha$ is a regularization parameter which needs to be tuned as below explained; and $S[\rho]$ stands for the Shannon-Jaynes entropy
\begin{equation}\label{eq:entropy}
S\left[\rho\right] = \int_0^{\infty} d\tilde{r}\, \left[\rho(\tilde{r})-\rho_0(\tilde{r})-\rho(\tilde{r}) \log \left(\frac{\rho(\tilde{r})}{\rho_0(\tilde{r})}\right)\right] \,,
\end{equation}
with $\rho_0$ denoting a prior estimate that, in the following, will be simply implemented by considering the well-known VMD model. Thus, given a form factor behaving as
\begin{equation}\label{eq:VMD}
F_0(Q^2)=\frac{M^2_{\rm{v}}}{M^{2}_{\rm{v}} +Q^2} \,,
\end{equation}
where $M_{\rm{v}}$ is vector meson's mass taken from Table~\ref{tab:mf}, the prior in Eq.\,\eqref{eq:entropy} will be replaced by 
\begin{subequations}
\label{eq:prior}
\begin{align}\label{eq:prior3d}
\rho_0^{\rm{3D}}(r) &= \frac{M^2_{\rm{v}}}{4\pi}\frac{e^{-M_{\rm{v}}r}}{r} \,, \\
\label{eq:prior2d}
\rho_0^{\rm{2D}}(b) &= \frac{M^2_{\rm{v}}}{2\pi}K_0(M_{\rm{v}}b) \,,
\end{align}
\end{subequations}
either in 3D or 2D cases, where $K_0$ is the zeroth modified Bessel function of the second kind.

As our input for $F(Q_i^2)$ comes from the calculations in the DSEs/BSEs framework described in the previous section, the same unique error will be assumed at any $Q_i$, \emph{i.e.}, $\sigma_i=\sigma$; and the parameter $\alpha$ is redefined: $\alpha \to \bar{\alpha}=\sigma^2 \alpha$\,\cite{Mueller:2010ah}. Then, we determine $\bar{\alpha}$ and the corresponding distribution function $\rho(\tilde{r})$ by applying cross-validation methods widely used in machine learning\,\footnote{Note here that, in the standard MEM~\cite{Asakawa:2000tr}, the best $\alpha$ and distribution functions $\rho(\tilde{r})$ are determined in terms of probability as long as $\sigma^2_i$ are provided. However, the uncertainty associated with numerical computations in DSEs/BSEs framework is difficult to be determined. Therefore, one can opt for estimating $\sigma^2_i$~\cite{Gao:2016jka} or comparing the results for different $\alpha$~\cite{Mueller:2010ah}, we choose the second one  supplemented with machine learning techniques in order to improve accurateness.}. Indeed, considering that we can produce sufficient data in the low-momentum region, $\sim \mathcal{O}(10^2)$, it works as follows: (i) one divides the ensemble of data for $F(Q_i^2)$ into two sets, a training one collecting a $70\%$ of data, $F(Q^2_k)$, and a testing one, $F(Q^2_j)$, made by the remaining $30\%$; (ii) solve then Eq.~\eqref{eq:MEM} based on the training set $F(Q^2_k)$, obtaining the distribution functions $\rho(\tilde{r},\bar{\alpha})$ and predicting therewith a value at any $Q_j^2$ for the test set; (iii) select the best $\bar{\alpha}$ as that generating the least error between the predicted and test values, \emph{viz.} Eq.\,\eqref{eq:error}; and, finally, (iv) repeat steps (i-iii) by collecting data for the training and test sets in different ways. As a balance of accuracy and efficiency, we chose to group all produced data in bins of 10 elements which, as explained above, can be partitioned each in $C(10,3)=120$ different ways.

\begin{figure}
\centering
\includegraphics[trim = 0mm 5mm 0mm 0mm, clip, width=1.0\columnwidth]{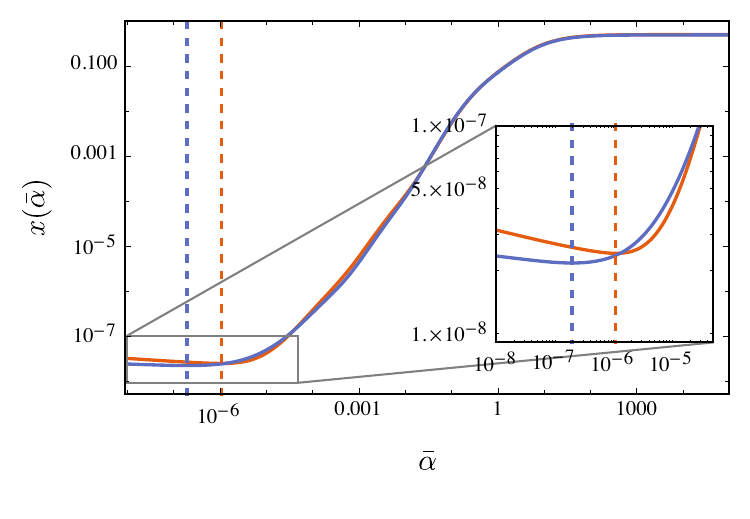}
\caption{\label{fig:x} $x(\bar{\alpha})$ defined by Eq.\,\eqref{eq:error}, expressing the error between model's prediction and test set, as a function of the parameter $\bar{\alpha}$, calculated for the reconstruction of the 3D charge distribution of the $\rho$ meson. The vertical dashed lines mark the minima resulting from distinct partitions of the data set (in different colors).}
\end{figure}

The error between the predicted and test values can be expressed as
\begin{equation}\label{eq:error}
x(\bar{\alpha}) = \sum_j \left[ F(Q_j^2) - \int_0^{\infty} d\tilde{r}\, \mathcal{K}(Q_j,\tilde{r}) \rho(\tilde{r},\bar{\alpha}) \right]^2,
\end{equation}
which appears displayed in Fig. \ref{fig:x} for two illustrative partitions of the data ensemble. It can be therein seen that, as $\bar{\alpha}$ gradually decreases, $x(\bar{\alpha})$ decreases first, reaches a minimum and increases beyond. This can be interpreted as owing to a model transitioning from under-fitting to over-fitting. The minimum of $x(\bar{\alpha})$ indicates the best $\bar{\alpha}$, which depends on the partition choice. We then average all the obtained $\rho(\tilde{r})$ from different partitioning of the data ensemble to determine the final distribution function, and estimate a statistical error band from the dispersion of results. Additionally, in order to check the sensitivity of the results to the prior estimate, we rescale the used VMD model, Eqs.~\eqref{eq:prior}, by 0.5-2 times and merge it into the error band~\cite{Qin:2014dqa}. A further scrutiny of the impact from the choice of the prior estimate is also performed, specifically for the lightest pseudoscalar and vector mesons, and described in App.\,\ref{ap:largeQ2}.

\section{NUMERICAL RESULTS}
\label{sec:Results}

\subsubsection{Charge distribution of charged mesons}

Within the DSEs/BSEs framework described in Sec.\,\ref{sec:FFs-DSEsBSEs}, we calculate first the electric form factor of pseudo-scalar and vector mesons in the light-quark sector, \emph{i.e.}, $\pi$, $K$, $\rho$, $K^*$ mesons; and extend further the calculation to pseudo-scalar mesons constituted from $q=u/d$-like quarks with current masses equal to those of the $s$, $c$ and $b$ quarks, namely, $\pi_s$, $\pi_c$, $\pi_b$, respectively, and their associated vector partners $\rho_{s}$, $\rho_c$, $\rho_b$. As masses and leptonic decay constants are concerned, the $\rho_{s,c,b}$ and $\pi_{c,b}$ results can be compared with those for $\phi$, $J/\psi$, $\Upsilon$, $\eta_c$ and $\eta_b$ mesons~\cite{Xu:2019ilh, Qin:2019oar, Xu:2021mju}. Then, the charge distributions are reconstructed based on the form factor's data obtained within the momentum domain $[0, Q^2_{\text{max}}]$, as illustrated in Table~\ref{tab:r}.

\begin{table}[!t]
\caption{\label{tab:r} Charge radii obtained from form factors (``Slope") and charge distributions (``Integration"). As a comparison, the experimental values of $\sqrt{\left<r^2\right>}$ are~\cite{ParticleDataGroup:2022pth, Povh:1990ad}: $\pi: 0.659(4)$, $K_\pm: 0.560(31)$,  $K_0: 0.277(18)i$ and $\rho: 0.721(35)$. The units are fm, except that for $Q^2_{\rm{max}}$ which, given in $\rm{GeV^2}$, represents the maximum value of transferred momentum used to reconstruct the charge distribution from the corresponding form factor. The errors quoted for the estimates from charge distributions only propagate the systematic uncertainty in MEM-based reconstruction as discussed in Sec.\,\ref{sec:Distributions}.}
\begin{ruledtabular}
\begin{tabular}{lllllc}
& \multicolumn{2}{c}{$\sqrt{\langle r^2 \rangle}$} & \multicolumn{2}{c}{$\sqrt{\langle b^2 \rangle}$} & \multicolumn{1}{c}{$Q^2_{\rm{max}}$} \\ 
& Slope & Integration & Slope & Integration & \\
\hline
$\pi$     & 0.646    & 0.646(1)    & 0.527    & 0.528(1)    & 2.2 \\
$K_\pm$   & 0.601    & 0.601(1)    & 0.491    & 0.491(1)    & 3.0 \\
$\pi_s$   & 0.466    & 0.466(1)    & 0.380    & 0.381(1)    & 4.0 \\
$\pi_c$   & 0.227    & 0.227(1)    & 0.185    & 0.185(1)    & 5.6 \\
$\pi_b$   & 0.113    & 0.113(1)    & 0.092    & 0.092(1)    & 8.5 \\
$\rho$    & 0.722    & 0.722(1)    & 0.589    & 0.590(1)    & 2.2 \\
$K^*_\pm$ & 0.645    & 0.646(1)    & 0.527    & 0.527(1)    & 2.2 \\
$\rho_s$  & 0.520    & 0.520(1)    & 0.424    & 0.425(1)    & 3.0 \\
$\rho_c$  & 0.247    & 0.247(1)    & 0.202    & 0.202(1)    & 3.2  \\
$\rho_b$  & 0.120    & 0.120(1)    & 0.098    & 0.098(1)    & 5.0 \\ \hline
$K_0$     & $0.284i$ & $0.283(1)i$ & $0.232i$ & $0.231(1)i$ & 3.0 \\
$K_0^*$   & $0.271i$ & $0.267(1)i$ & $0.221i$ & $0.218(1)i$ & 2.2 \\
\end{tabular}
\end{ruledtabular}
\end{table}

As discussed in Sec.~\ref{sec:Distributions}, the charge radius can be computed either directly from the derivative of the corresponding form factor at vanishing momentum, or from the first non-trivial moment of the charge distributions (\emph{viz.}, Eqs.\,\eqref{eq:r2n} and \eqref{eq:b2n} for $n$=1),  
\begin{subequations}
\label{eq:ri}
\begin{align}
\langle{r^2}\rangle &= \int_0^{\infty} dr\, r^2\ 4\pi r^2\rho^{\rm{3D}}(r) \,, \\
\langle{b^2}\rangle &= \int_0^{\infty} db\, b^2\ 2\pi b\rho^{\rm{2D}}(b) \,.
\end{align}
\end{subequations}
The quality of the charge distributions derived by inversion of Eq.\,\eqref{eq:FTgen} within MEM framework, can be first evaluated through a comparison of the charge radii obtained from their integration in Eqs.\,\eqref{eq:ri} and obtained from the form factors, \emph{viz.} Eqs.\,(\ref{eq:3dr2},\ref{eq:2db2}). This comparison is shown in Tab.\,\ref{tab:r}, and exhibits that the two determinations are plainly consistent. However, as illustrated by Fig.~\ref{fig:x}, radii are quite insensitive to the pointwise behavior of the charge distribution, precisely because they are defined in terms of the particular distribution moment expressing the average of $r^2$, which is linked to a local property of the form factor by a mathematical identity.  All together, two main observations can be extracted from Tab.\,\ref{tab:r}: (i) the charge radius of charged mesons decreases as the current-quark masses increase; and (ii) the one for every vector meson is always larger than its pseudo-scalar partner's. 

Aiming at completion, the values obtained from applying Eqs.\,\eqref{eq:ri} for the neutral mesons $K_0$ and $K_0^\ast$ have been incorporated into Tab.\,\ref{tab:r}, although their case will be discussed in the next subsection. At this point, we will only underline that their form factors are endowed with a positive derivative at zero momentum, entailing a negative $\langle r^2 \rangle$ and, hence, an imaginary radius. Despite that unphysical outcome, Eqs.\,(\ref{eq:3dr2},\ref{eq:2db2}) are still valid and the determinations from both the form factor and the charge distribution are consistent. Indeed, Eqs.\,\eqref{eq:ri} implies that a negative $\langle r^2 \rangle$ relies on a  positive non-definite distribution charge.
\begin{figure*}[!t]
\centering\includegraphics[trim = 0mm 8mm 0mm 0mm, clip, width=1.0\columnwidth]{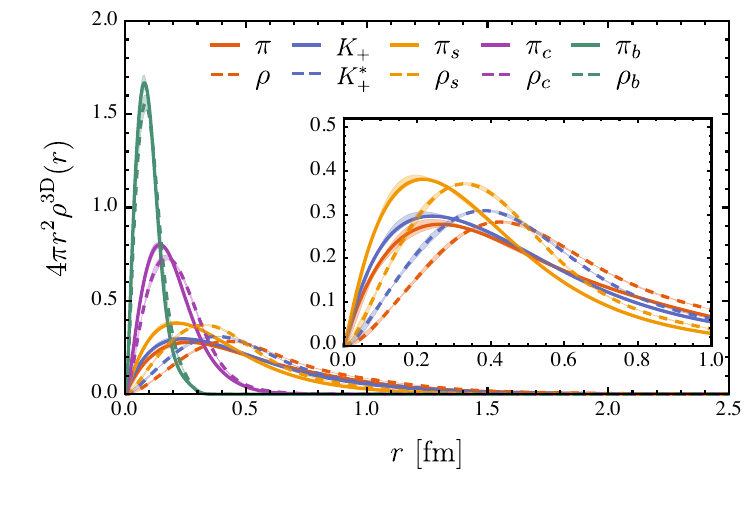}
\centering\includegraphics[trim = 0mm 8mm 0mm 0mm, clip, width=1.0\columnwidth]{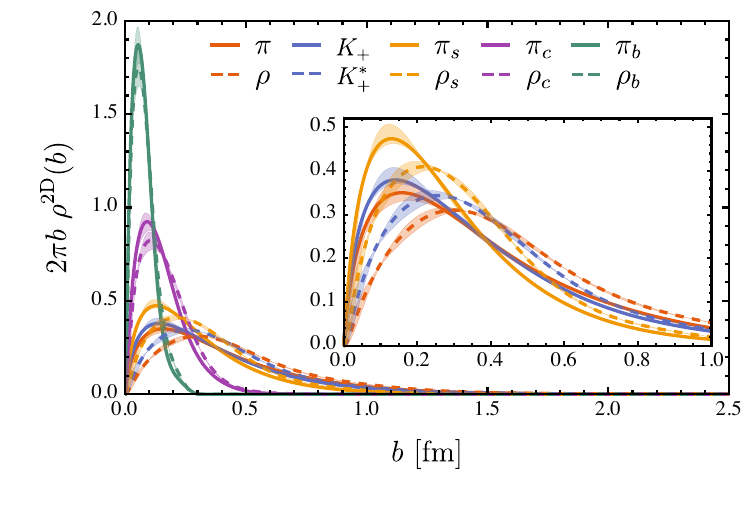}
\centering\includegraphics[trim = 0mm 8mm 0mm 0mm, clip, width=1.0\columnwidth]{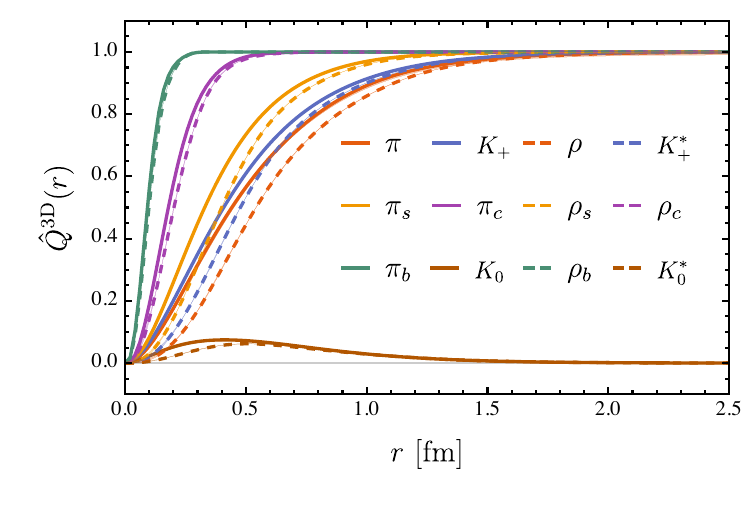}
\centering\includegraphics[trim = 0mm 8mm 0mm 0mm, clip, width=1.0\columnwidth]{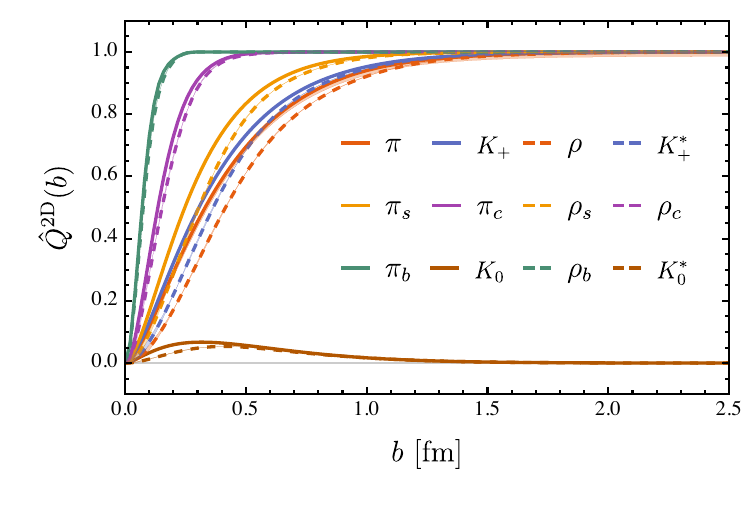}
\caption{\label{fig:charged} \emph{Upper panels:} 3D and 2D charge distribution of pseudo-scalar mesons and vector mesons; \emph{Lower panels:} Charge covered by different radius as explained in the text and Eqs.~\eqref{eq:ChargeCovered}. Solid lines: pseudo-scalar mesons, dashed lines: vector mesons.}
\end{figure*}

Furthermore, the integration of charge distributions below a given radius, 
\begin{subequations}
\label{eq:ChargeCovered}
\begin{align}
\widehat{Q}^{\rm{3D}}(r_\textrm{max}) &=\int_{0}^{r_\textrm{max}} dr\, 4\pi r^2 \rho^{\rm{3D}}(r) \,, \\
\hat{Q}^{\rm{2D}}(b_\textrm{max}) &= \int_{0}^{b_\textrm{max}} db\, 2\pi b \,\rho^{\rm{2D}}(b) \,,
\end{align}
\end{subequations}
define a quantity that can be classically interpreted as a charge covered up to that radius, and which offers a further insight into the matter of the spatial extension of the system. Obviously, the covered charge amounts to the total charge, given by normalization in Eq.\,\eqref{eq:norm}, as the radius approaches infinity. In Table~\ref{tab:covered}, we present the radii corresponding to a covered charge equal to 50\%, 70\% and 90\% of the total charge.

\begin{table*}[!t]
\caption{\label{tab:covered} Radii corresponding to different covered charges, the unit is fm.}
\begin{ruledtabular}
\begin{tabular}{lllllllll}
&\multicolumn{4}{c}{3D} &\multicolumn{4}{c}{2D}\\ 
&\multicolumn{1}{c}{50\%}& \multicolumn{1}{c}{70\%} & \multicolumn{1}{c}{90\%}&\multicolumn{1}{c}{$\sqrt{\langle r^2\rangle}$} &\multicolumn{1}{c}{50\%}& \multicolumn{1}{c}{70\%} & \multicolumn{1}{c}{90\%}&\multicolumn{1}{c}{$\sqrt{\langle b^2\rangle}$}\\
\hline
$\pi$& 0.442(5) & 0.646(6) & 1.025(15)& 0.646(1)& 0.331(7) & 0.506(8) & 0.850(23) & 0.528(1)\\
$K_\pm$& 0.408(2) & 0.592(3) & 0.954(6)&0.601(1) &0.306(3) & 0.466(3) &0.787(6) & 0.491(1)\\
$\pi_s$& 0.330(1) & 0.469(1) & 0.735(1)& 0.466(1)&0.248(1) & 0.369(1) & 0.609(1)&0.381(1) \\
$\pi_c$& 0.179(1) & 0.239(1) & 0.346(1)&0.227(1) &0.136(1) & 0.190(1) & 0.290(1)&0.185(1)\\
$\pi_b$& 0.093(1) & 0.120(1) & 0.168(1)&0.113(1) &0.071(1) & 0.096(1) & 0.141(1)& 0.092(1)\\
$\rho$ & 0.542(1) & 0.732(1) & 1.113(1)&0.722(1) &0.416(2) & 0.584(2) & 0.927(2)&0.590(1)\\
$K^*_\pm$& 0.477(1) & 0.643(1) & 0.998(1)&0.646(1) &0.365(2)&0.514(1) &0.824(1)& 0.527(1)\\
$\rho_s$& 0.400(1) & 0.532(1) & 0.797(1)&0.520(1) &0.305(2) &0.427(1) &0.662(1)&0.425(1)\\
$\rho_c$& 0.203(1) & 0.264(1) & 0.369(1)&0.247(1) & 0.153(1) &0.212(1) &0.312(1)& 0.202(1)\\
$\rho_b$& 0.099(1) & 0.128(1) & 0.180(1)&0.120(1) &0.075(1) &0.103(1) &0.153(1)& 0.098(1)\\
\end{tabular}
\end{ruledtabular}
\end{table*}

\begin{figure*}[!t]
\centering\includegraphics[trim = 0mm 8mm 0mm 0mm, clip, width=1\columnwidth]{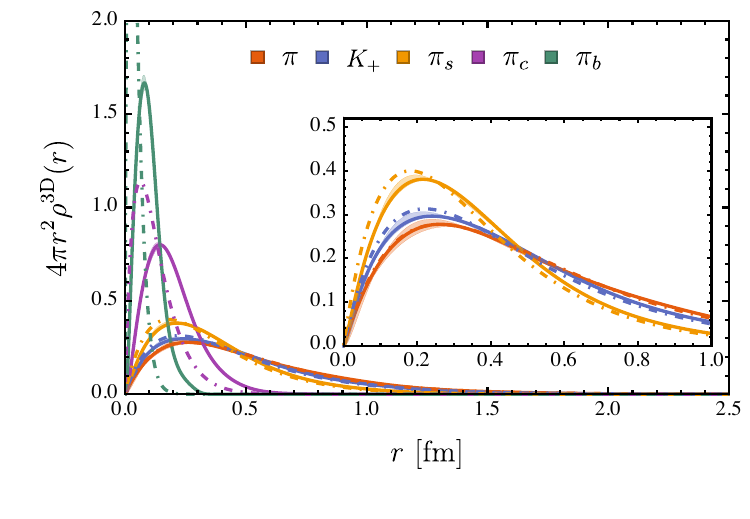}
\centering\includegraphics[trim = 0mm 8mm 0mm 0mm, clip, width=1\columnwidth]{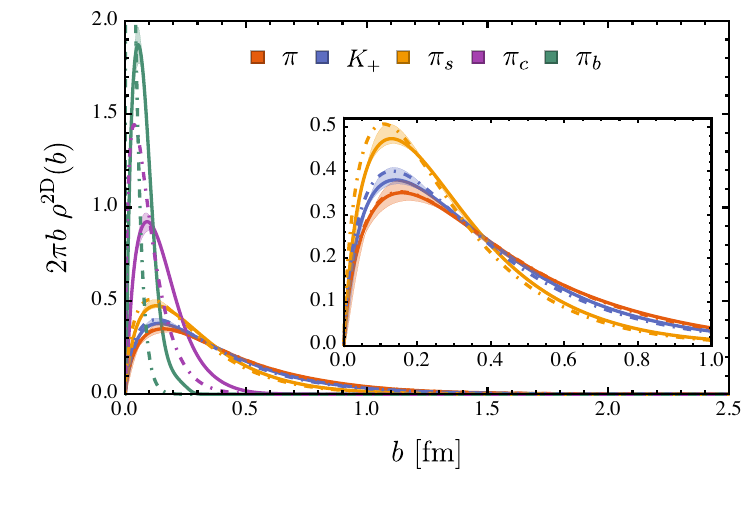}
\centering\includegraphics[trim = 0mm 8mm 0mm 0mm, clip, width=1.0\columnwidth]{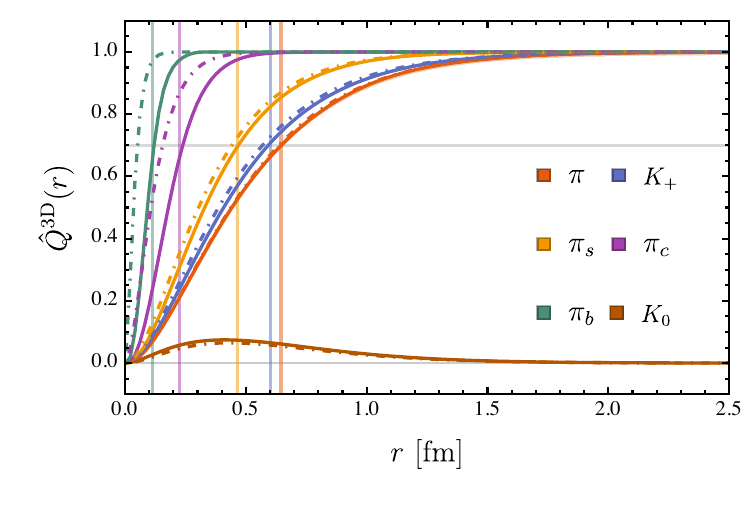}
\centering\includegraphics[trim = 0mm 8mm 0mm 0mm, clip, width=1.0\columnwidth]{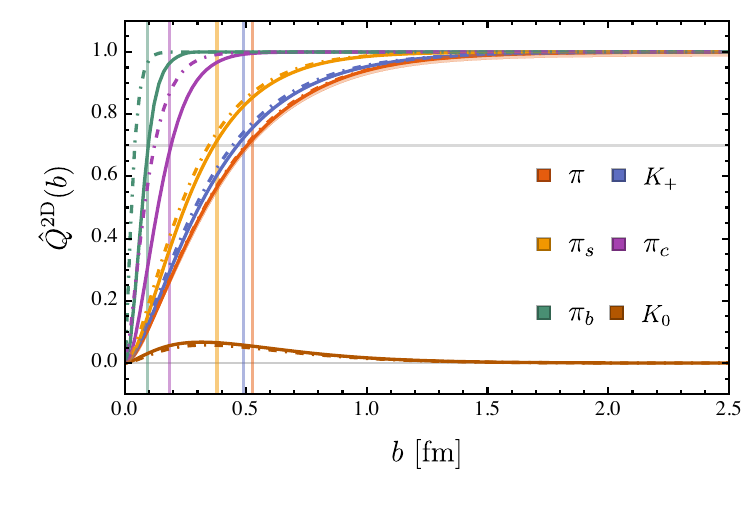}
\caption{\label{fig:pscomp} Comparison between the results of DSEs/BSEs and VMD model. Solid lines: DSEs/BSEs, dot-dashed lines: VMD. The vertical lines indicates the charge radii obtained by DSEs/BSEs.}
\end{figure*}

\begin{figure*}[!t]
\centering\includegraphics[trim = 0mm 8mm 0mm 0mm, clip, width=1\columnwidth]{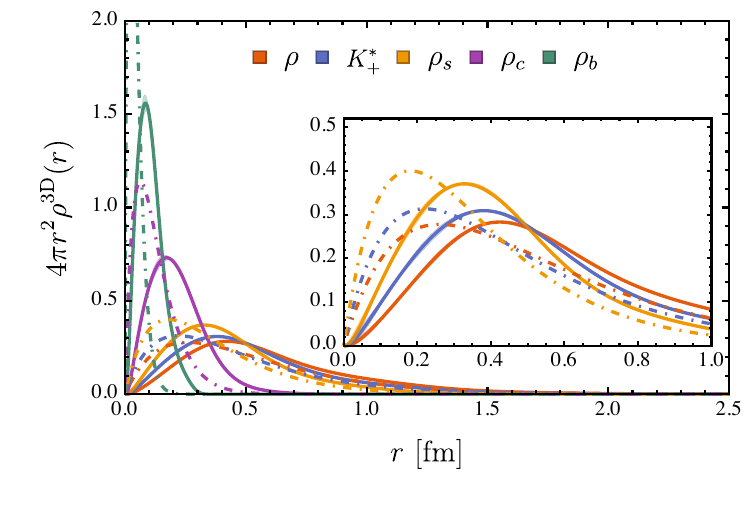}
\centering\includegraphics[trim = 0mm 8mm 0mm 0mm, clip, width=1\columnwidth]{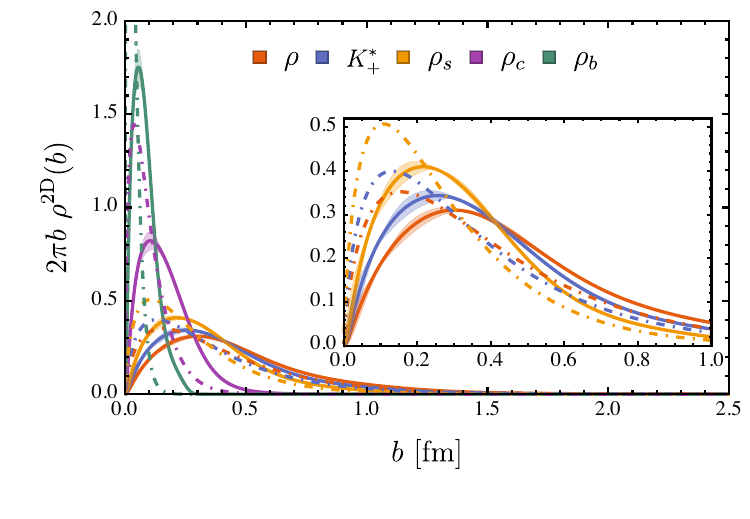}
\centering\includegraphics[trim = 0mm 8mm 0mm 0mm, clip, width=1.0\columnwidth]{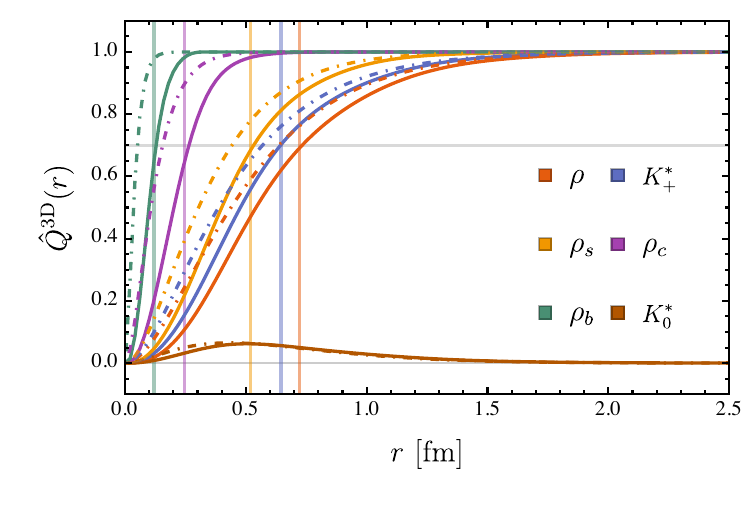}
\centering\includegraphics[trim = 0mm 8mm 0mm 0mm, clip, width=1.0\columnwidth]{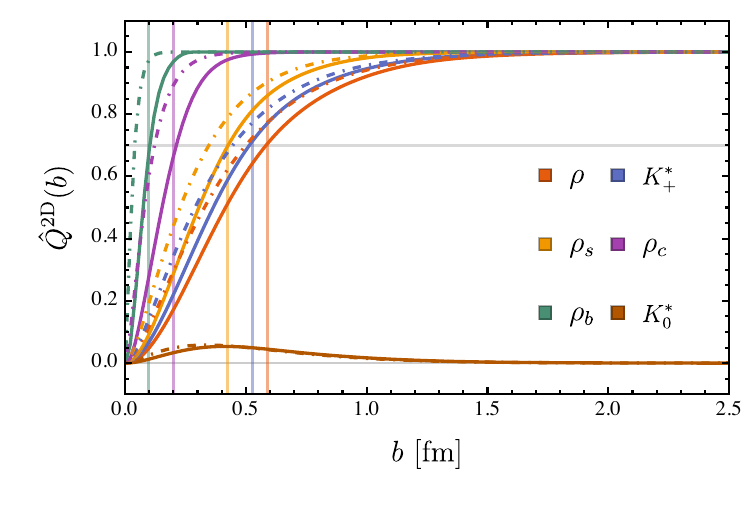}
\caption{\label{fig:vccomp} Comparison between the results of DSEs/BSEs and VMD model. Solid lines: DSEs/BSEs, dot-dashed lines: VMD. The vertical lines indicates the charge radii obtained by DSEs/BSEs.}
\end{figure*}

\begin{figure}[!t]
\centering\includegraphics[trim = 0mm 15mm 0mm 0mm, clip, width=1.0\columnwidth]{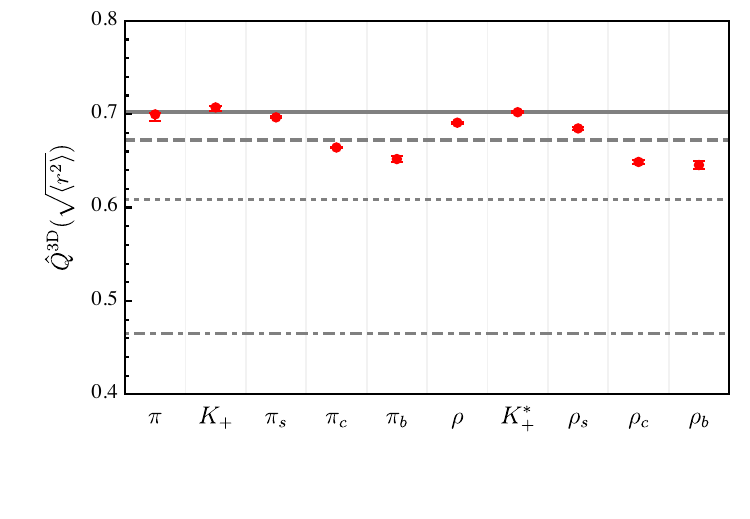}
\caption{\label{fig:type} The charge covered by $\sqrt{\langle r\rangle^2}$ in four typical 3D distributions and their comparison with our results (red points). Solid line: Yukawa distribution; dashed line: exponential distribution; dotted line: Gaussian distribution; dot-dashed line: hard-sphere distribution.}
\end{figure}

The reconstructed 3D and 2D charge distributions for charged mesons appear displayed, respectively, in the left- and right-upper panels of Fig.~\ref{fig:charged}. Their comparison makes clearly apparent that 2D distributions are more compressed near the center than 3D, due to the reduction of the canvassed dimensions, although the shape and the pattern for the different mesons that each shows, as expected, remain the same. 

Concerning this last pattern, as the meson system transitions from light to heavy quark sectors, the charge distribution tends to shrink, which basically means that the spatial range of motion is larger for a light than for a heavy quark. In the case that only the lightest quark flavors are involved ($\pi$, $\rho$ and $K_+$ and $K_+^\ast$), the charge density is still significantly non-zero at $r$ or $b \sim 1$ fm; while it is zero already at around 0.6 fm when the current quark masses are increased to reach that of the charm ($\pi_c$ and $\rho_c$), and at around 0.4 fm when the bottom quark mass is taken ($\pi_b$ and $\rho_b$). When comparing vector and pseudoscalar partners, the former's spatial extension is significantly larger than the latter's; their difference being maximal in the lightest sector and diminishing gradually for heavier mesons, becoming negligible when bottom quark masses are involved. This appears to suggest that spin-dependent interactions become less significant when current masses of the meson's constituent quarks increase, and the Higgs mechanism weights more than DCSB in the meson's mass budget. 

All the latter is consistent with the radii hierarchy apparent in Tab.\,\ref{tab:r}, and with the conclusions we extracted from. Equivalently, the same can be inferred from the evolution of the covered charge with the radius of integration, shown in the lower panels in Fig.\,\ref{fig:charged}, or from the outputs of Tab.\,\ref{tab:covered}. In the lower panels, apart from charged mesons' results, those for neutral $K_0$ and $K_0^\ast$ are also included, and shown to exhibit a different behavior which stems from their positive non-definite charge distribution and their null total charge. The discussion of these features is postponed for the next subsection. 

In the aim of providing more elements of comparison, Figs.~\ref{fig:pscomp} (pseudoscalar mesons) and \ref{fig:vccomp} (vector) display the reconstructed charge distributions (upper panels) and covered charge (lower) together with the predictions from VMD model. Clearly, VMD and DSEs/BSEs results for light pseudo-scalar mesons are compatible, but they significantly differ in all vector cases and for the heavy mesons herein considered. The difference, again, gradually grows up with the current quark mass. 

We furthermore illustrate our results by a comparison of the covered charge at the standard radius, $\hat{Q}^{\rm{3D}}(\sqrt{\langle r^2 \rangle})$, obtained with our 3D charge distributions and with other four typical ones: Yukawa, exponential, Gaussian and hard-sphere distributions; usually considered for pions, protons, light nuclei and heavy nuclei, respectively~\cite{Kumano:2017lhr}. Deviations are apparent for heavy mesons, our estimations indicating less spatial extension and lying below (above) the prediction from exponential (Gaussian) distributions.     


\subsubsection{Charge distribution of neutral mesons}

\begin{figure}[!t]
\centering\includegraphics[trim = 0mm 5mm 0mm 0mm, clip, width=1.0\columnwidth]{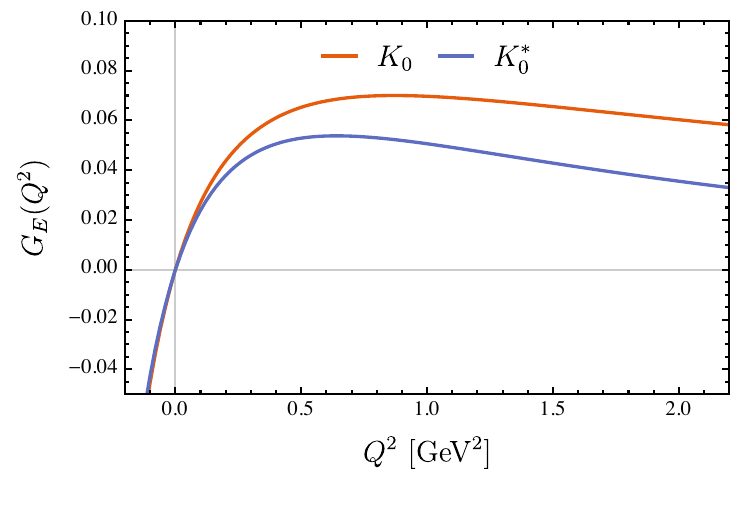}
\caption{\label{fig:kff} Electric form factor of $K_0$ and $K_0^*$ mesons.}
\end{figure} 

\begin{figure*}[!t]
\centering\includegraphics[trim = 0mm 5mm 0mm 0mm, clip, width=1.0\columnwidth]{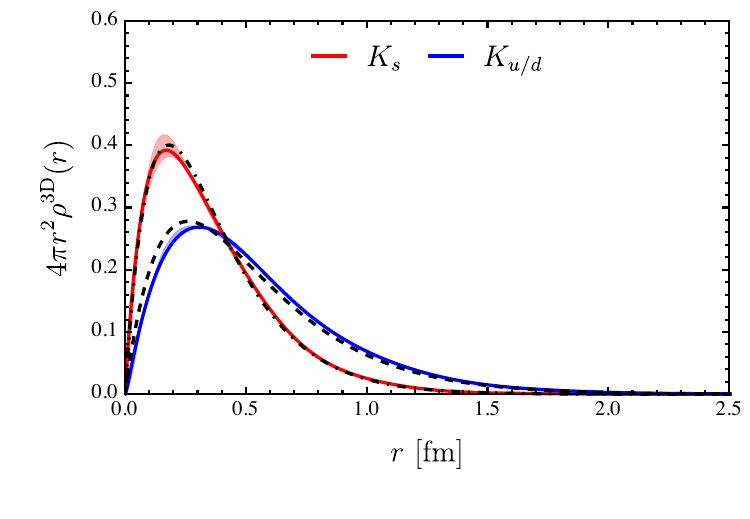}
\centering\includegraphics[trim = 0mm 5mm 0mm 0mm, clip, width=1.0\columnwidth]{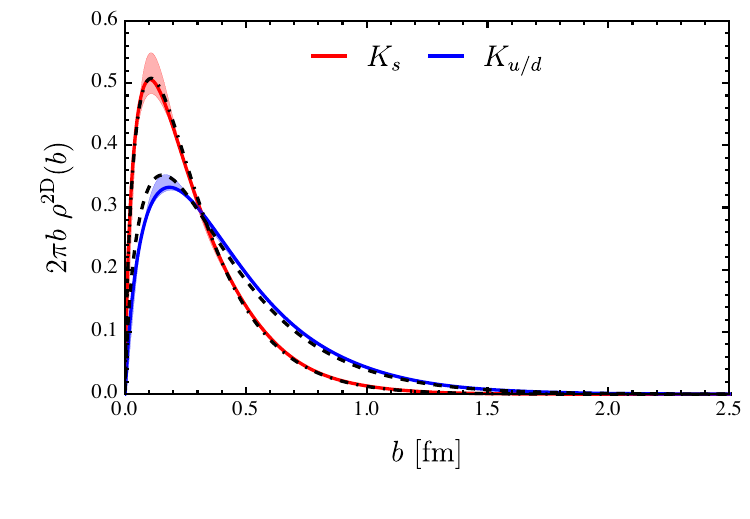}
\centering\includegraphics[trim = 0mm 5mm 0mm 0mm, clip, width=1.0\columnwidth]{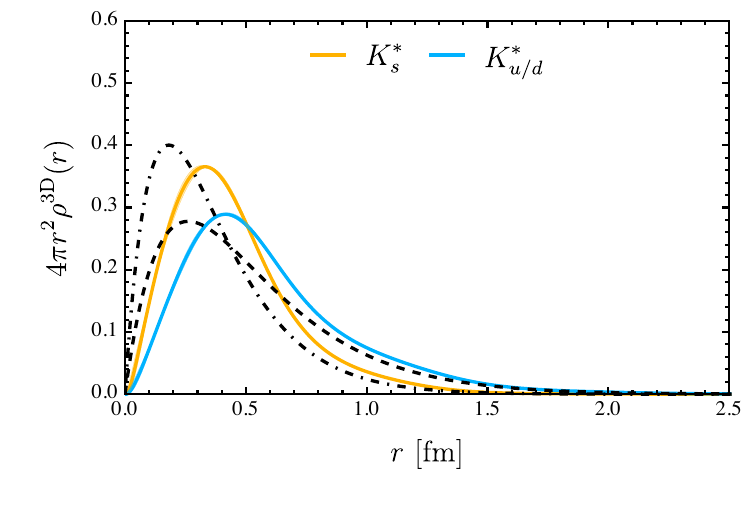}
\centering\includegraphics[trim = 0mm 5mm 0mm 0mm, clip, width=1.0\columnwidth]{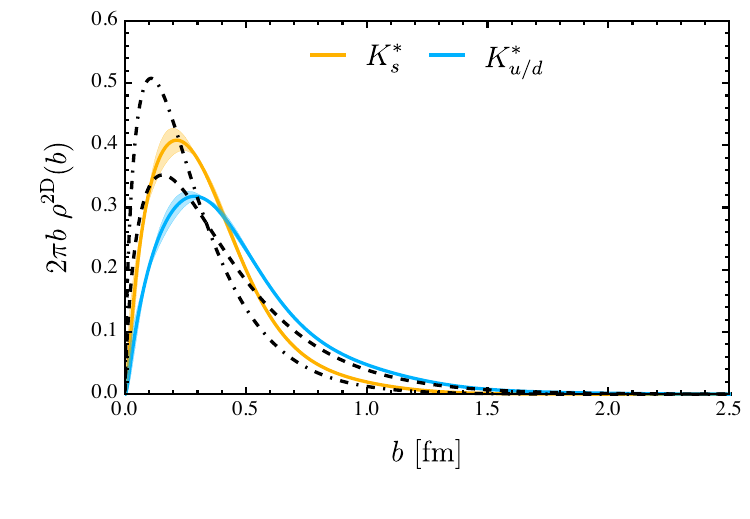}
\centering\includegraphics[trim = 0mm 5mm 0mm 0mm, clip, width=1.0\columnwidth]{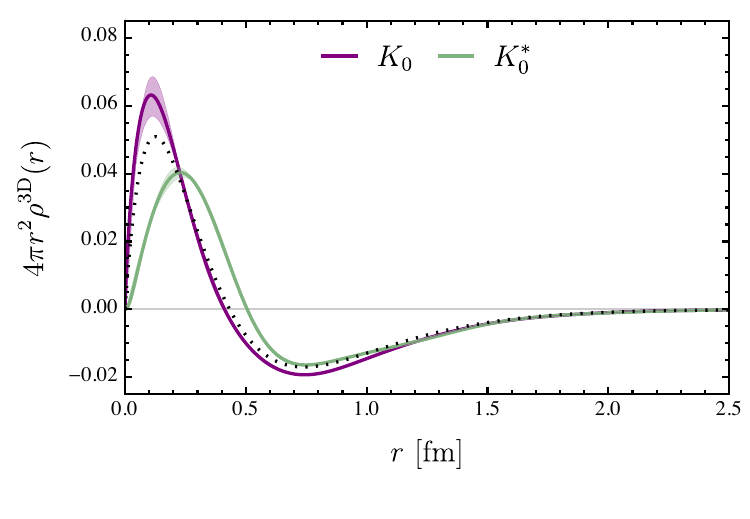}
\centering\includegraphics[trim = 0mm 5mm 0mm 0mm, clip, width=1.0\columnwidth]{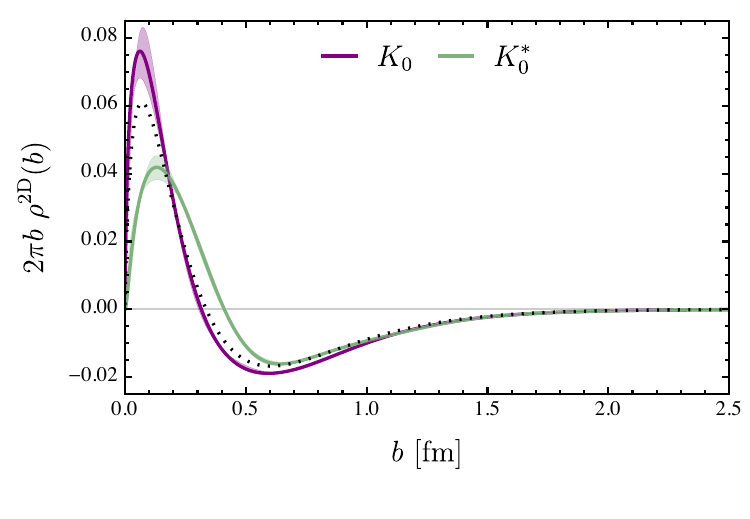}
\caption{\label{fig:k0} The 3D and 2D charge distributions of the $K_0$ and $K^*_0$ mesons. Solid lines: DSEs/BSEs; dashed lines: VMD of $K_{u/d}(K_{u/d}^*)$; dot-dashed lines: VMD of $K_s(K_s^*)$; dotted lines: VMD of $K_0(K_0^*)$. \emph{Upper panels:} charge distributions of $d$ and $s$ quark in $K_0$ meson; \emph{middle panels:} charge distributions of $d$ and $s$ quark in $K^*_0$ meson; \emph{lower panels:} the full charge distribution of $K_0$ and $K_0^*$ meson.}
\end{figure*}

In the case of the neutral $K_0$ and $K_0^*$ mesons, their size cannot be expressed in terms of a charge radius given by Eqs.\,\eqref{eq:ri}. Their form factors possess a positive derivative at vanishing momentum (see Fig.\,\ref{fig:kff}) and, consequently, $\langle r^2 \rangle$ becomes negative, implying that their associated charge distributions are positive non-definite\,\footnote{This also causes some numerical instabilities in the reconstruction of the charge distribution by MEM.}. Indeed, the meson form factors are the result of combining two quark contributions which, in the neutral case, produce a destructive interference by their opposite charge. Namely,  
\begin{align}
F^{K_0}(Q^2) = -\frac{1}{3} F^{K_0}_d(Q^2) + \frac{1}{3} F^{K_0}_{\bar{s}}(Q^2) \,,
\end{align}
and analogously for $K^\ast_0$. Therefore, owing to the linearity of \eqref{eq:FTgen}, a charge distribution can be separately reconstructed from each quark contribution, and both combined as 
\begin{align}\label{eq:rhosum}
\rho^{K_0}(\tilde{r}) = -\frac{1}{3} \rho^{K_0}_d(\tilde{r}) + \frac{1}{3} \rho^{K_0}_{\bar{s}}({\tilde{r}}) \,.
\end{align} 
Thus, each expresses a spatial distribution for a total charge amounting to a positive unity and, weighted by their own charge, they overlap to produce the full meson's charge distribution. This destructive interplay of two opposite contributions explains that, in the lower panel of Figs.\,\ref{fig:charged}, \ref{fig:pscomp} and \ref{fig:vccomp}, the peak of covered charge for the neutral mesons is more than one order of magnitude smaller than the saturation value for charged mesons. 

For MEM reconstruction, the VMD model is again used as prior with $\rho$ and $\rho_s$ masses for $d$ and $s$ cases, respectively, thus implementing Eqs.\,(\ref{eq:VMD},\ref{eq:prior}). The outputs for the reconstructed charge distribution are presented in Fig.~\ref{fig:k0}. The same pattern shown for charged mesons can be seen here: 2D distributions (right panels) more compressed near the origin than 3D (left). Here, the quark contributions for $K_0$ (upper panels) and $K_0^\ast$ (middle) are separately displayed and, in both cases, the lighter-quark density extends farther than the heavier. This can be intuitively interpreted as the consequence of a heavier quark weighting more for the definition of the meson's center of momentum and, hence, lying closer to this center than the lighter. Then, both densities combine to give the full distributions shown in the lower panels:  near the center, positive charge dominates the interplay and, as the probe gradually moves away from it, the dominance is reversed, the meson's charge distribution becoming negative after crossing a zero. 

Again, the pseudoscalar system appears to be more compact than the vector. Instead of the (negative) average of $r^2$, one can use the position of the zero crossing to evaluate and compare the neutral meson's sizes.  Table~\ref{tab:k0} shows these zeroes: $r_0$ (3D) and $b_0$ (2D). As can be see, $K^*_0$'s is about $(20-30)\%$ larger than $K_0$'s. Our estimates compare fairly well to VMD predictions in the pseudoscalar case, but significantly differ for the vector. This is also consistent with the analysis of charged mesons.

\begin{table}
\caption{\label{tab:k0} Neutral radius, in fm, of $K_0$ and $K_0^*$ meson. See text for details.}
\begin{ruledtabular}
\begin{tabular}{ccccc}
& \multicolumn{2}{c}{$r_0$ [fm]} & \multicolumn{2}{c}{$b_0$ [fm]} \\ 
& DSEs/BSEs & VMD & DSEs/BSEs & VMD  \\
\hline
$K_0$    & $0.412(7)$ & $0.433$  & $0.316(9)$ & $0.336$ \\
$K_0^*$  & $0.505(1)$ & $0.433$  & $0.411(5)$ & $0.336$ 
\end{tabular}
\end{ruledtabular}
\end{table}


\section{SUMMARY}
\label{sec:Summary}

Aiming at delivering charge distributions of pseudoscalar and vector mesons, both in spatial 3D and transverse 2D, we investigated their electromagnetic form factors within a DSE/BSE framework which constrains the outputs below a maximum transferred momentum of $\sim$ 2 GeV in most of the cases. 
Consequently, as form factors and charge distributions are related by a Fourier transform which, \emph{a priori}, requires being known over the full support, DSE/BSE results need to be supplemented with a reconstruction procedure. To this purpose, we have used the maximum entropy method, supplied by a Bayesian prior fixed by a VMD model with a vector meson mass given by the corresponding homogeneous BSE. Although the VMD model only accounts for low-momentum DSE/BSE results for the lightest pseudoscalar case (and does not incorporate scaling violations), the sensitivity of MEM outputs for charge distributions to prior's choice has been canvassed and their reliability proved down to distances of 0.1 fm. Interestingly, supplementing the DSE/BSE results with our proposed MEM reconstruction, a differential feature for vectors and pseudoscalar mesons can be predicted and interpreted; namely, the form factor exhibition of a zero crossing in the vector channel.  

The overall analysis shows that the charge distribution of meson gradually shrinks around the center of (transverse) momentum in 3D (2D) with the increase of the current-quark mass. Moreover, the charge distribution of a vector meson extends always farther than that of its pseudo-scalar partner, the difference decreasing gradually when transitioning from light to heavy quark sectors. The 2D distributions are more compressed than 3D, but their shapes and trends with quark-current mass and meson's properties are shown to be the same. Concerning the size, the standard squared average (\emph{viz.}, Eq.\,\eqref{eq:3dr2}) is shown to cover about 70\% of the total charge for the light mesons, which agrees with the expectation for a Yukawa distribution; but the quark-current and resulting meson masses slightly lower this percentage, which takes around 65\%, when $c$ or $b$ flavors are involved, a value lying between the expectations from exponential and Gaussian distributions.  

In the case of neutral $K_0$ and $K_0^*$ mesons, they both exhibits a non-definite positive charge distribution, resulting from the non-trivial interplay of those for their quark constituents, the one corresponding to the $s$-quark being more compact than that associated with the $u/d$-quark. The consequence is a charge density which is positive near the center of momentum, crosses zero and becomes then negative; where the peaks are about one order of magnitudes smaller than those for charged mesons. Moreover, this zero crossing could be taken as a measure of the size, at least for comparative purposes, of the neutral mesons. Thus, our numerical results suggest that $K^*_0$ is about $(20-30)\%$ larger than $K_0$.

Finally, as an immediate perspective of this work, the reconstruction method herein used to extract charge distributions from electromagnetic form factors could be also considered to derive mass distributions from gravitational form factors. 


\begin{acknowledgments}
This work has been partially funded by Ministerio Espa\~nol de Ciencia e Innovaci\'on under grant Nos. PID2019-107844GB-C22 and PID2022-140440NB-C22; Junta de Andaluc\'ia under contract Nos. Operativo FEDER Andaluc\'ia 2014-2020 UHU-1264517 and PC-I+D+i under the title: "\emph{Tecnologías avanzadas para la exploración del universo y sus componentes}" (Code AST22-0001). The authors acknowledge, too, the use of the computer facilities of C3UPO at the Universidad Pablo de Olavide, de Sevilla.
\end{acknowledgments}

\appendix 

\section{The convergence of Eq.\,\eqref{eq:Fmom}}
\label{app:convergence}

After Taylor's expansion of $j_0$ in Eq.\,\eqref{eq:3d}, the exchange of integral and sum entailing Eq.\,\eqref{eq:Fmom} can be only allowed when the corresponding series, before and after the exchange, are proved to be convergent. The former is guaranteed, given the radius of convergence for the series from the Taylor's expansion of $j_0$; while, for the latter,  
$\rho^\textrm{3D}(r)$ needs to be assumed to decrease faster than any power of $r$, when $r$ goes to infinity, so that $\langle r^{2n}\rangle$ defined by Eq.\,\eqref{eq:r2n} is always a non-singular c-number. 

Furthermore, although the form factor in Eq.\,\eqref{eq:3d} might be well defined for all $Q^2$, the domain of convergence for the series in Eq.\,\eqref{eq:Fmom} depends on the particular charge distribution and could be restricted to a given range of $Q^2$. Notwithstanding this, Eq.\,\eqref{eq:Fmom} can be always approached as an asymptotic expansion and, as such, delivering reliable results. Namely, given a few number $m$ of distribution moments $\langle r^{2n} \rangle$, there will always be a $Q_0^2$ such that, for any $Q^2 \le Q_0^2$, the series in Eq.\,\eqref{eq:Fmom}'s rhs can be truncated for $n=0,\dots, m$; delivering therewith a good approximation for the form factor in lhs. Particularly, one can take derivatives and specialize at $Q^2=0$, 
\begin{align}\label{eq:A1}
\left. \frac{d^n}{d^nQ^2} F(Q^2) \right|_{Q^2=0} = (-1)^n \frac{n!}{(2n+1)!} \langle r^{2n} \rangle\;,     
\end{align}
relating thus moments distributions and form factor derivatives. 

Let us illustrate all the latter by considering the VMD model, Eqs.\,(\ref{eq:VMD},\ref{eq:prior3d}). Then, Eq.\,\eqref{eq:r2n} reads
\begin{align}\label{eq:A2}
\langle r^{2n} \rangle = \frac 1 {M_\nu^{2n}} \int_0^\infty dx \, x^{2n+1} e^{-x} = \frac{\Gamma(2n+2)}{M_\nu^{2n}} = \frac{(2n+1)!}{M_\nu^{2n}} \;,  
\end{align}
which, plugged into Eq.\,\eqref{eq:Fmom} gives
\begin{align}\label{eq:A3}
F(Q^2) = 1 + \sum_{n=1}^\infty (-1)^n \left(\frac{Q^2}{M_\nu^2}\right)^n \;;  
\end{align}
entailing that the convergence of the series is restricted to $Q^2 < M^2_\nu$. Clearly, Eq.\,\eqref{eq:A3} corresponds to the Taylor expansion of $F_0(q^2)$ from Eq.\,\eqref{eq:VMD}. Taking derivatives in both sides and applying Eq.\,\eqref{eq:A1}, one can equivalently recover \eqref{eq:A2}; \emph{e.g.}, $\langle r^2 \rangle = 6/M_\nu^2$.   

\section{Sensitivity to prior's functional behavior}
\label{ap:largeQ2}

As explained in Sec.\,\ref{sec:Distributions}, the inversion of Eq.\,\eqref{eq:FTgen} \emph{via} MEM etails the optimization of \eqref{eq:MEM} with the likelihood function \eqref{eq:Lrho} and the Shannon-Jaynes entropy \eqref{eq:entropy}. The latter requires the implementation of a (bayesian) prior, herein taken from a VMD model entirely defined by a vector meson mass derived from the corresponding homogeneous BSE. The initial prior's large-momentum behavior is made to evolve through MEM optimization and deliver thus a final MEM solution which, at low momenta, is attached to input data (from DSE/BSE form factors) by the likelihood function. As far as the low-distance profile of the charge density relies on the form factor large-momentum behavior, it is pertinent questioning the impact of prior's on MEM solution's behavior and, hence, on the low-distance profile. 

\begin{figure*}[!t]
\centering\includegraphics[trim = 0mm 5mm 0mm 0mm, clip, width=1.0\columnwidth]{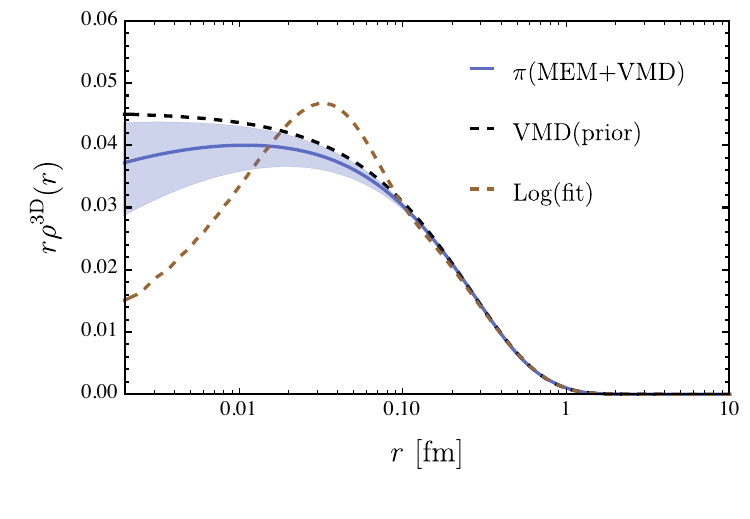}
\centering\includegraphics[trim = 0mm 5mm 0mm 0mm, clip, width=1.0\columnwidth]{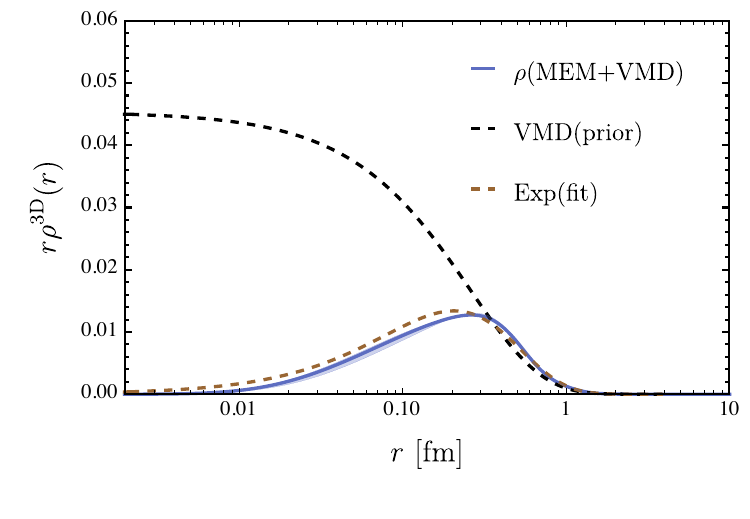}
\centering\includegraphics[trim = 0mm 5mm 0mm 0mm, clip, width=1.0\columnwidth]{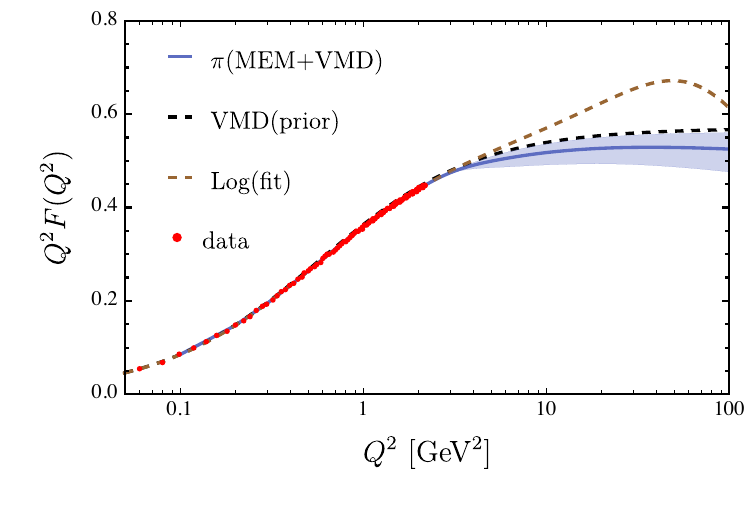}
\centering\includegraphics[trim = 0mm 5mm 0mm 0mm, clip, width=1.0\columnwidth]{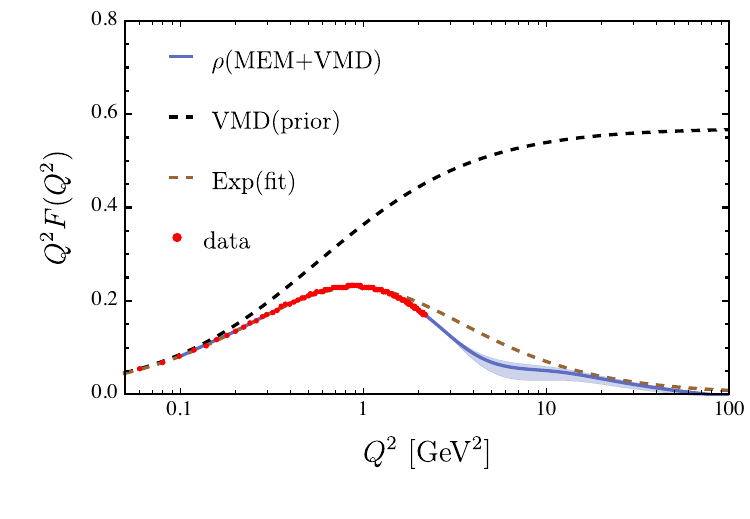}
\caption{\label{fig:add} The upper panels show $r \rho^\textrm{3D}(r)$ for the lightest pseudoscalar (left) and vector (right) mesons, while the lower ones display the corresponding form factors, $Q^2 F(Q^2)$. The red solid circles stand for the DSE/BSE calculation of the form factors (lower) which are numerically inverted by the implementation of MEM with a VMD prior (displayed in black dashed line) to give the charge distribution in blue solid line (upper). Form factors are reobtained by applying Eq.\,\eqref{eq:3d} and displayed also in blue solid lines (lower). The blue bands express the systematic uncertainty from the numerical inversion. For the sake of comparison, accurate fits to the DSE/BSE data and their corresponding inverted charge distributions are shown with a brown dashed line in, respectively, lower and upper panels.
}
\end{figure*}

We can elaborate further on this issue by the inspection of the panels of Fig.\,\ref{fig:add}, displaying $r\rho^\textrm{3D}(r)$ (upper) and $Q^2 F(Q^2)$ (lower) for the lightest pseudoscalar (left) and vector (right) mesons. The VMD prior, a particular fit and the MEM results appear therein compared for both cases. We focus on the 3D case (analogous arguments works also for 2D) and choose to display $r\rho^\textrm{3D}(r)$ because Eq.\,\eqref{eq:3d} can be recast as follows
\begin{align}\label{eq:Fpos}
F(Q^2) =& \frac{4\pi}{Q} \sum_{n=0}^\infty \int_0^{\pi/Q} 
dr \sin{(Qr)}  \\
&\times \left[ \phi\left(r+2n\frac{\pi}{Q}\right) - \phi\left(r+(2n+1)\frac{\pi}{Q}\right) \right] \;,   
\nonumber
\end{align}
with $\phi^\textrm{3D}(r)=r \rho^\textrm{3D}(r)$. Then, Eq.\,\eqref{eq:Fpos} makes clearly apparent that a necessary condition for a form factor exhibiting a zero crossing and sign reversing is its delivering a charge distribution such that $r \rho^\textrm{3D}(r)$ does not monotonically decreases with $r$. Otherwise, all the terms of the sum in Eq.\,\eqref{eq:Fpos} would be positive definite. On the other hand, as far as $\rho^\textrm{3D}(r)$ decreases with $r$ faster than $1/r$ in the large-distance domain, one can then naturally conclude that $r \rho^\textrm{3D}(r)$ is suppressed at $r$=0 and, in the low-momentum domain, increases with $r$. 

A zero crossing is known to happen for $F(Q^2)=G_E^{VC}(Q^2)$, in the vector case. This could be physically interpreted as the consequence of a spin-dependent contribution to the quark-antiquark effective interaction, which is expected to separate away spin-aligned quarks (vector) and draw near spin-anti-aligned ones (pseudoscalar). Consequently, a larger support for the charge density at small distances (near the center of momentum) is expected for the latter case in respect to the former. Investigating whether the MEM analysis can capture this differential feature for pseudoscalar and vectors from low-momentum data from DSE/BSE deserves therefore attention. 

In the pseudoscalar case, the VMD form factor reproduces very well the DSE/BSE data and, at large momenta, decreases as $F(Q^2) \sim Q^{-2}$; while the fitting expression, following Ref.\,\cite{Xu:2023izo} (\emph{viz.} Eq.\,(27) therein), keeps an accurate description of data and behaves as $F(Q^2) \sim Q^{-2}/\ln{Q^2}$ at large momenta (incorporating a logarithmic suppression accounting for the scaling violations expected for a 4D quantum field theory). Concerning the vector case, VMD does not account for DSE/BSE data and, for comparative purposes, we have considered as fitting expression 
\begin{align}\label{eq:exp}
F(Q^2) = \frac{1}{(1+ a Q^2)^2} \;\; \Rightarrow \;\; \rho^\textrm{3D}(r) = \frac 1 {8\pi a^3} e^{-r/a} \;,   
\end{align}
which, with $a=1.046$, reproduces fairly well the data up to $Q^2 \simeq$ 2 GeV$^2$.

In the upper panels of Fig.\,\ref{fig:add}, $r \rho^\textrm{3D}(r)$ for the VMD model is shown as a monotonically decreasing function. In the pseudoscalar case,  it is in good agreement with the MEM result in the whole range, and only differ below 0.1 fm from the numerical inversion of \eqref{eq:3d} with the best fit to DSE/BSE data as input\footnote{If, in applying MEM, one replaces the VMD prior with this numerical inversion of the best fit, the MEM solution would be shown to be plainly consistent with the inversion.}. In the vector case, both the MEM solution (with VMD prior) and the numerical inversion of \eqref{eq:3d} with the best fit of Eq.\,\eqref{eq:exp} to DSE/BSE data are consistent with each other and strongly differ from the VMD prior. In the lower panels, both for pseudoscalar and vector, $Q^2 F(Q^2)$ obtained from the VMD prior, the MEM solution and the best fits are displayed and compared to DSE/BSE data.  Clearly, the likelihood contribution to \eqref{eq:MEM} makes, in both cases, the optimal MEM solution to be constrained by the DSE/BSE input and, irrespectively of the prior, this is good enough to deliver a reliable charge density down to very low distances. Specially in the vector case, the presence of a zero crossing is captured by the MEM solution and shown to correspond with the congruent low-distance density behavior, despite the implementation of a prior lacking it.

\bibliography{ChargeDistributions.bib}

\end{document}